\documentclass[12pt]{article}
\usepackage{graphicx}

\newcommand\pubdate{\today}
\def\kpnn{K^{+}\rightarrow\pi^{+}\nu\bar{\nu}}
\def\support{\footnote{for the NA62 Collaboration:

G.~Aglieri Rinella, R.~Aliberti, F.~Ambrosino, B.~Angelucci, A.~Antonelli, G.~Anzivino, 

R.~Arcidiacono, I.~Azhinenko, S.~Balev, J.~Bendotti
% ...
}}

\def\Title#1{\begin{center} {\Large #1 } \end{center}}
\def\Author#1{\begin{center}{ \sc #1} \end{center}}
\def\Address#1{\begin{center}{ \it #1} \end{center}}

\newcommand\pubblock{\rightline{\begin{tabular}{l} \\
         \pubdate  \end{tabular}}}
\newenvironment{Abstract}{\begin{quotation}  }{\end{quotation}}
\newenvironment{Presented}{\begin{quotation} \begin{center} 
             PRESENTED AT\end{center}\bigskip 
      \begin{center}\begin{large}}{\end{large}\end{center} \end{quotation}}

\def\beq{\begin{equation}}
\def\eeq#1{\label{#1}\end{equation}}
\def\eeqn{\end{equation}}

\def\beqa{\begin{eqnarray}}
\def\eeqa#1{\label{#1}\end{eqnarray}}
\def\eeqan{\end{eqnarray}}

\let\bar=\overbar

\def\Dslash{\not{\hbox{\kern-4pt $D$}}}
\def\dslash{\not{\hbox{\kern-2pt $\del$}}}

\def\msb{{\bar{\ssstyle M \kern -1pt S}}}

\begin{document}
\begin{titlepage}
\pubblock

\vfill
\Title{The NA62 experiment at CERN:
status and perspectives}
\vfill
\Author{ Riccardo Fantechi\support}
\Address{INFN Sezione di Pisa and CERN}
\vfill
\begin{Abstract}
The NA62 experiment at CERN will measure the branching ratio of the $\kpnn$ decay with a precision of 10\%, collecting about 100 events in two years of run. This paper will review the design of the experiment, the strategy for the measurement and the actual status of the installation. Prospects for the data collection and for the measurement of other rare $K^+$ channels will be presented.
\end{Abstract}
\vfill
\begin{Presented}
$12^{th}$ Flavour Physics and CP Violation\\
Marseille, June $26^{th}-30^{th}$, 2014 
\end{Presented}
\vfill
\end{titlepage}
\def\thefootnote{\fnsymbol{footnote}}
\setcounter{footnote}{0}

\section{Introduction}
The ultra rare decays $K \rightarrow \pi\nu\bar{\nu}$ are, among the many rare flavour changing neutral
current K and B decays, very interesting in the
 search for new physics through underlying mechanisms of flavour
mixing. Several circumstances allow to compute the SM branching ratio to very high precision:
\begin{itemize}
\item{the $O(G^2_F)$ electroweak amplitudes exhibit a power-like GIM mechanism}
\item{the top-quark loops largely dominate the matrix element} 
\item{the sub-leading charm-quark contributions have been computed at NNLO order \cite{NNLO}}
\item{the hadronic matrix element can be extracted from the branching ratio of the
$K^+ \rightarrow \pi^0 e^+\nu$ decay, well known experimentally \cite{hadronic}}.
\end{itemize}
The latest prediction for the $\kpnn$ channel
is ($7.81 \pm 0.75 \pm 0.29) \cdot 10^{−11}$ \cite{brod}. The first error comes from the uncertainty on the CKM matrix
elements, the second one is the pure theoretical uncertainty. This decay is one of the best probes
for new physics effects complementary to the LHC, especially within non Minimal Flavour Violation
models \cite{isidori}\cite{blanke}. Since the extreme theoretical cleanness of these decays remains also in these
scenarios, even deviations from the SM value at the level of 20\% can be considered signals of new
physics. Also, the decay can be used for a measurement of $V_{td}$ free from hadronic uncertainties
and independent from that obtained with B mesons decays. Nevertheless only a measurement of the branching ratio with at least 10\% accuracy can be a significant test of new physics.
 
The decay $\kpnn$ has been
observed by the experiments E787 and E949 at the Brookhaven National Laboratory and the measured
branching ratio is $1.73^{+1.15}_{-1.05}\cdot10^{-10}$ \cite{e787}. The KOTO experiment at J-PARC is searching for the complementary $K^0 \rightarrow \pi^0 \nu \bar{\nu}$ decay and will present during the summer a preliminary result from a short run.

The NA62 experiment at CERN-SPS \cite{na62} will measure the branching ratio of $\kpnn$ with a 10\% error.

\section{Goals of the NA62 experiment}

NA62 aims to collect about 100 $\kpnn$ decays in two years of running, in order to measure the branching ratio with 10\% precision, with an acceptance of $\sim$ 10\%. This measurement requires first of all a beam line capable to provide at least $10^{13}$ kaon decays, integrated over the data taking period. In addition
a background rejection factor O($10^{12}$) is required, to extract the signal in the overwhelming abundance of the various frequent decay modes of $K^+$. Indeed $K^+ \rightarrow \pi^+\pi^0$ (21\%) and $K^+ \rightarrow \mu^+ \nu$ (63\%) should be reduced using respectively photon veto detectors and muon identifiers. The use of a high energy kaon beam (75 GeV/c) has an advantage in the first case, because, imposing a cut on 
$P_{\pi^+}<35$ GeV/c, the energy of the photons from the $\pi^0$ is larger than 40 GeV and the efficiency of a photon detector is very high. Additional particle ID is needed to eliminate other backgrounds and to identify the parent kaon inside the unseparated beam.

The beam line is the same used by NA62 for the $R_K$ measurement \cite{rk}, which can provide the required intensity of about $3\cdot10^{12}$ protons/pulse at an energy of 75 GeV with 1\% momentum bite. The beam contains $\pi^+$ (70\%), protons (23\%) and $K^+$ (6\%). The rate seen by the detectors along the beam line, integrated over a surface of 12.5 cm$^2$, is about 750 MHz. The decay region is defined to be 80 m of decay volume after the last beam line element. The decay volume is evacuated down to $10^{-6}$ mbar, to minimize the multiple scattering of the decay products and the number of interactions of the beam with the residual gas. The downstream detectors are designed to measure the decay products in this volume. The rate downstream is given mainly by kaon decays and is about 10 MHz.

\begin{figure}[htb]
\centering
\includegraphics[height=3in]{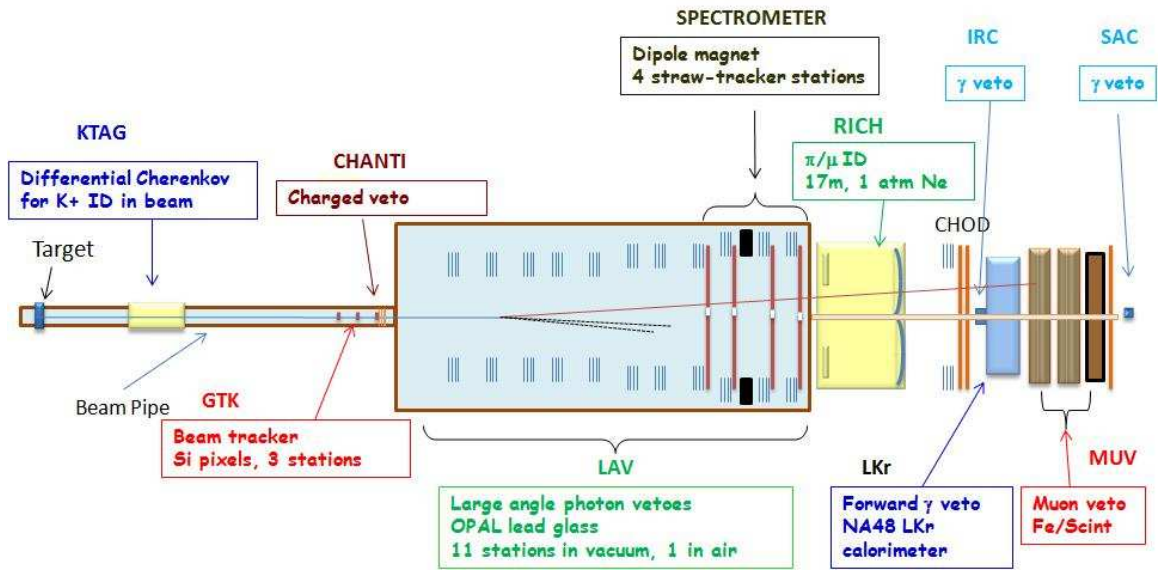}
\caption{Layout of the NA62 detector.}
\label{fig:layout}
\end{figure}

The layout of the detector is shown in fig.\ref{fig:layout}; the functions of the individual detectors will be described below.

\section{Strategy of the measurement}

\begin{figure}[htb]
\centering
\includegraphics[height=2in]{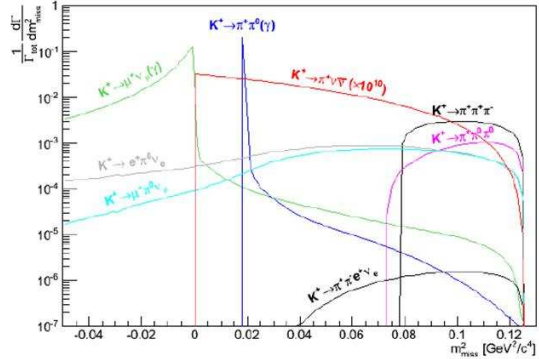}
\caption{Missing mass distribution for signal and background channels. Background channels are normalized to the branching ratios. The signal distribution is multiplied by $10^{10}$.}
\label{fig:mm2}
\end{figure}

The signature of the signal events is one incoming kaon and one charged output track. The missing mass squared $m^2_{miss}=(P_K-P_{\pi})^2$  is a first handle to eliminate the background. In fig.\ref{fig:mm2} the missing mass distribution for the signal and the various background channels is shown. In addition, accidental signal track matched with a K-like one can mimic background contributions.

Two regions in the squared missing mass spectrum, on both sides of the $\pi^0$ peak, are chosen to select candidate events: Region 1 extends from 0 to 0.01 GeV$^2$/c$^4$ and 
region 2 from 0.026 to 0.068 GeV$^2$/c$^4$.

The experiment needs good tracking devices to measure both four momentum of the parent kaon and the pion of the decay. The good performances of the tracking devices allow to use the missing mass cut with a rejection of $O(10^5)$ for the $K^+ \rightarrow \mu^+ \nu$ ($K_{\mu2}$) decay 
and of $O(10^4)$ for the $K^+ \rightarrow \pi^+ \pi^0$ ($K_{\pi2}$).

Veto detectors are needed to reduce further the background of the $K_{\mu2}$ and $K_{\pi2}$ decay channels.
The $K_{\mu2}$ background is reduced by a factor $O(10^5)$, while photon veto detectors reduce the $K_{\pi2}$
background by a factor $O(10^8)$.

An additional requirement is a good particle identification, both in the beam line, to identify a kaon in a beam composed mainly by pions, and in the decay volume, to detect positrons and to have an additional muon identification needed to have an additional $O(10^2)$ reduction of the $K_{\mu2}$.

Last but not the least, a precise timing is needed to match incoming kaons with the decay products.

\subsection{Reconstruction of the kaon beam}

The four momentum of the incoming kaon is measured using the Gigatracker beam spectrometer. It is composed of three silicon pixel stations, placed in vacuum, mounted respectively before, in between and after a system of achromat magnets.
The pixels of the Gigatracker stations are $300 \cdot 300 \mu$m$^2$, the material budget across the beam is less than 0.5\% $X_0$ per station, given by the sensor thickness ($200 \mu$m), the read-out chips 
($100 \mu$m) and a micro-channel cooling plate for the cooling of the detector. An important characteristic of the detector is the capability to handle a beam rate of 750 MHz. An extensive beam test of Gigatracker prototypes has been performed \cite{gtk}, giving the following performances:
\begin{itemize}
\item{$\sigma(t) \sim $ 200 ps/station}
\item{$\sigma(P_K)/P_K = 0.2\%$}
\item{$\sigma(dX,Y/dZ)_K = 15 \mu $rad}
\end{itemize}

To reduce the critical background from inelastic interactions in it, the last Gigatracker station is complemented by six stations of counters called CHANTI, each of which is composed of four layers of triangular shaped bars readout by WLS fibers and silicon photomultipliers.

\subsection{Reconstruction of the charged pion}

The four momenta of the charged decay products are measured using a spectrometer composed of 4 straw chambers and one dipole magnet. To minimize multiple scattering, the straw chambers are built with ultra-light material and are installed in the vacuum tank \cite{straw}. Each chamber has 4 views (x, y, u, v), each with 4 planes of staggered straw tubes (9.8 mm diameter, 2.1 m long), for a total of 7168 straws. The material seen by the particles is $X/X_0 \approx 0.1\%$ per view, $\rightarrow 2\%$ in total. The dipole magnet has a $P_t$ kick of 265 MeV/c. The expected resolutions are:
\begin{itemize}
\item{$\sigma(P)/P = (0.32 \oplus 0.008 P)\%$, with P in GeV/c}
\item{$\sigma(dX,Y/dZ) = 20-50 \mu$rad}
\end{itemize}

\subsection{Particle identification}
\subsubsection{Kaon identification}
The parent kaon in the beam is identified by the KTAG, a Cherenkov detector based on the existing SPS differential Cherenkov (CEDAR) device, which has been upgraded to sustain the 50 MHz kaon rate with an additional external optics projecting out the eight light spots from the CEDAR over 48 photomultipliers in
each spot. The CEDAR detector can be 
filled with nitrogen or with hydrogen, the former suiting at best the dispersion characteristics of the existing optics, the latter giving less material on the beam line.

The KTAG has been partially commissioned with Nitrogen in Autumn 2012, with a time resolution of less  than 100 ps and a sub-percent pion mis-tagging for a kaon efficiency of $> 95$\%.
\subsubsection{Photon detection}

The background of events with photons in the final state should be reduced by a factor at least $10^8$. It is possible to achieve this result with a system of photon detectors which is hermetic up to 50 mrad: only 0.2\% photons from $K+ \rightarrow \pi^+ \pi^0$ escape at an angle larger than 50 mrad and have anyway a very low energy and a very high detection inefficiency, but at the same time the other photon in the decay, with high energy, is detected with high efficiency in one of the forward detectors.

The photon veto system is composed of three detector types:
\begin{itemize}
\item{Twelve large angle photon veto (LAV) stations distributed along the decay volume, covering the
angular region from 50 down toto 8.5 mrad, built with a creative reuse of the lead glass blocks of the former OPAL electromagnetic calorimeter; the blocks are mounted in a 4 or 5 rings, with a small staggering to have a complete hermeticity. Photon detection inefficiency is better than $10^{-4}$ down to 100 MeV}

\item{The NA48 Liquid Krypton electromagnetic calorimeter \cite{na48lkr}, whose inefficiency has been  measured from data to be better than $10^{-5}$ for energy larger than 3 GeV, covering from 1 to 8.5 mrad. An upgrade of the readout electronics, needed to cope with the high trigger rate, is being commissioned now}

\item{Two Shashlik calorimeters covering down to 0 mrad, one (IRC) in front of the LKr calorimeter inner flange, the other (SAC) at the end of the hall, just before the beam dump.}
\end{itemize}
\subsubsection{Pion-muon separation}

A first pion-muon rejection capability is done using the Muon veto (MUV) detectors. Their functionality is twofold: the first two measure, as hadronic calorimeters,  the energy and the shower shapes of the incident particles, complementing the good capabilities of  the LKr calorimeter, while the third is a fast plane of scintillator designed to provide a fast signal to be used as an anticounter in the L0 trigger.

The first two muon detectors (MUV1, MUV2) are built with a sandwich of iron and scintillator, read with photomultipliers. The third detector (MUV3), installed after the other two and an iron wall,  is composed of $\sim 150$ 5 cm-thick scintillator tiles, each one read directly by two photomultipliers. This configuration allows to have fast signals and an inefficiency below 1\%.

The global pion-muon separation obtainable with these three detectors is of the order of $10^5$, which is still not enough to reach the needed value of $10^{12}$.

An additional rejection factor is then obtained with a RICH detector placed in front of the LKr calorimeter, 
after the last chamber of the spectrometer, to provide identification of a $\mu$  in the background and of a $\pi$ in the signal events. The RICH detector is $\sim 18$ m long, contains Neon at atmospheric pressure and has an array of 20 mirrors which focus the Cherenkov rings onto two spots (on the left and on the right of the beam pipe) with $\sim 1000$ photomultiplier each. The RICH vessel is traversed by a 9 cm radius beam pipe which transports the undecayed beam. The energy threshold for pions is $\sim 13$ GeV.

The RICH concept has been tested extensively in 2009 with a full length prototype \cite{rich}. The average muon misidentification has been measured to be 0.7\%; with improved analysis techniques it is possible to reduce this number to 0.5\% with a slight increase in the pion inefficiency. The RICH will also be used to provide a fast charged pion signal for the L0 trigger: during the test beam, the time resolution of the RICH has been measured to be less than 80 ps.

\subsection{Trigger and DAQ}

The high decay rate requires a powerful trigger system in order to reduce the frequency of signal candidates to few KHz. This is achieved with a multilevel trigger system \cite{trigger}, with a typical reduction of a factor 10 for each level. The first level (L0) is implemented directly in hardware, using the digitized time data from the RICH, the MUV3, the calorimeter and some of the large angle veto to construct digital "primitives" which are sent to a central L0 Trigger Processor to be combined and to produce a L0 trigger decision. The primitive rate is O(10 MHz) and the reduction factor in the L0TP is a factor 10. Data are then read from all the detectors but the calorimeter and sent to a farm of PCs where software L1 algorithms are run to reduce again the rate by a factor 10. On a positive L1 trigger, the calorimeter is read and further processing occurs on the global event, to reduce again by a factor 10 before sending data as a 10 KHz stream to tape.

\section{Signal-background ratio}
Extensive simulation work as well as the analysis of the data collected in the technical run in Autumn 2012 have been used to evaluate with a cut and count technique the level of the background expected in the final analysis.

\begin{figure}[htb]
\centering
\includegraphics[height=2in]{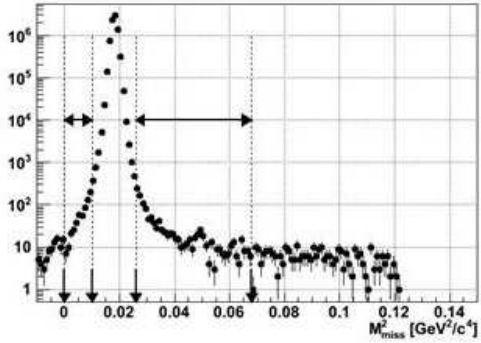}
\caption{Distribution of reconstructed $m^2_{miss}$ for $K^+ \rightarrow \pi^+ \pi^0$. The effect of the pileup is not included.}
\label{fig:mmass2pi0}
\end{figure}

The $K^+ \rightarrow \pi^+ \pi^0$ background is rejected by the photon vetoes and the missing mass cut in the two regions defined above. The resolution on $m^2_{miss}$ is $\approx 10^{-3}$, which allows to push the signal region towards the $\pi^0$ peak. However, as shown in fig. \ref{fig:mmass2pi0}, the non gaussian tails due to the multiple scattering and the pileup in the Gigatracker (inducing a possible wrong assignment of the measured $K^+$ track with the $\pi$ downstream) limit the kinematical rejection, worsening the $m^2_{miss}$ resolution by a factor 3. The simulation gives a value for the rejection of $5 \cdot 10^3$. Improvements are possible using the precise timing of Gigatracker, RICH and KTAG.

\begin{figure}[htb]
\centering
\includegraphics[height=2in]{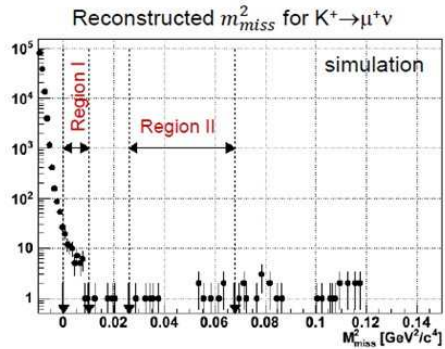}
\caption{Distribution of reconstructed $m^2_{miss}$ for $K^+ \rightarrow \mu^+ \nu$. The effect of the pileup is not included.}
\label{fig:mmass2mu}
\end{figure}

In the case of the decay $K^+ \rightarrow \mu^+ \nu$, the missing mass spectrum is negative and the cuts on $m^2_{miss}$ suppress strongly this type of background. Here again non gaussian tails and pileup (fig. \ref{fig:mmass2mu}) affect the rejection factor, which is estimated to be $1.5 \cdot 10^4$.

For the $K^+ \rightarrow \pi^+ \pi^+ \pi^-$ the requirement of only one track reconstructed in the spectrometer rejects $\approx 99$\% of the decays. The $m^2_{miss}$ cut is crucial to rejects events with a $\pi^+$ and a $\pi^-$ not reconstructed. The rejection factor from the kinematics is $1.5 \cdot 10^6$, again due to non gaussian tails in multiple scattering. The remaining events are rejected using combinations of cuts on the downstream detectors.

The main source of the accidental background is the interaction of the beam with the material along the beam line, namely the Gigatracker and the residual gas in the vacuum region. For its rejection, the KTAG performances (see above) are critical to ensure the reduction of about 94\% of the interactions of the $\pi^+$ and protons of the beam. 

The summary of the estimation of the signal to background ratio, using a cut and count analysis without any optimization, is shown in tab. \ref{tab:bkg}. The number of events is normalized to the $4.5 \cdot 10^{12}$ expected number of kaon decays per year. The Standard Model branching ratio for $\kpnn$ is assumed.

\begin{table}[h!]
\begin{center}
\begin{tabular}{l|c}  
Decay &  events/year   \\ \hline
 $\kpnn (SM) Flux=4.5 \cdot 10^{12}$     &     45    \\ \hline
 $K+ \rightarrow \pi^+ \pi^0$     &     5    \\ 
 $K+ \rightarrow \mu^+ \nu$     &     1    \\ 
 $K+ \rightarrow \pi^+  \pi^+ \pi^-$     &     $< 1 $   \\ 
 $K+ \rightarrow \pi^+ \pi^- e^+ \nu$ + other 3 track decays   &  $< 1$    \\ 
 $K+ \rightarrow \pi^+ \pi^0 \gamma (IB)$     &     1.5    \\ 
 $K+ \rightarrow \mu^+ \nu \gamma (IB)$     &     0.5    \\ 
 $K+ \rightarrow \pi^0 e^+ (\mu^+) \nu$, others     &     negligible    \\ \hline 
Total Background & $<10$ \\
\end{tabular} 
\caption{Expected signal and background events}
\label{tab:bkg} 
\end{center}
\end{table}

\section{Additional physics program}

The high beam rate and the powerful trigger system will allow to collect in parallel to the main decay data from many rare kaon decays, reaching unprecedented sensitivity. Table \ref{tab:addphy} show what will be the
reach of NA62. For example, in the dilepton decay modes one could improve by an order of magnitude the present limits. For the $R_K$ measurement the expected improvement of a factor 2 of the NA62 Phase 1 result is achieved not only from an increase in statistics, but also from a reduction in the systematic error due to the better detector. The same consideration is valid for the ChPT channels.
\begin{table}[h!]
\begin{center}
\begin{tabular}{c|c|c|c}  
     Decay&      Physics   &      Present limit  &     NA62 \\ \hline
     $\pi^+\mu^+e^-$   &   LFV   &  $1.3 \cdot 10^{-11}$   &  $0.7 \cdot 10^{-12}$ \\
     $\pi^+\mu^-e^+$   &   LFV   &  $5.2\cdot10^{-10}$   &  $0.7\cdot10^{-12}$ \\
     $\pi^-\mu^+e^+$   &   LNV   &  $5.0\cdot10^{-10}$   &  $0.7\cdot10^{-12}$ \\
     $\pi^-e^+e^+$     &   LNV   &  $6.4\cdot10^{-10}$   &  $2.0\cdot10^{-12}$ \\
     $\pi^-\mu^+e^+$   &   LNV   &  $1.1\cdot10^{-9}$   &   $0.4\cdot10^{-12}$ \\
     $\mu^-\nu e^+e^+$  & LFV/LNV   & $2\cdot10^{-8}$   &    $4.0\cdot10^{-12}$ \\
     $e^-\nu\mu^+\mu^+$   &   LNV   &  No data   &   $1.0\cdot10^{-12}$ \\ \hline 
     $\pi^+ \chi^0$   &      New particle   & $5.9\cdot10^{-11}$, $M\chi = 0 $  &     $1.0\cdot10^{-12}$ \\
     $\pi^+\chi\chi$   &      New particle   &      No data   &   $1.0\cdot10^{-12}$ \\
     $\pi^+\pi^+e^-\nu$   & $\Delta S \neq \Delta Q$   &$1.2\cdot10^{-8}$   &   $1.0\cdot10^{-11}$ \\
     $\pi^+\pi^+\mu^-\nu$   & $\Delta S \neq \Delta Q$    &$3.0\cdot10^{-6}$   &   $1.0\cdot10^{-11}$ \\
     $\pi^+\gamma$   &     Angular  momentum   &  $2.3\cdot10^{-9}$   &   $1.0\cdot10^{-11}$ \\
     $\mu^+\nu_h, \nu_h \rightarrow \nu \gamma$   &   Heavy neutrino  &   Limits up to $M\nu_h = 350 MeV/c^2$  & $1.0\cdot10^{-12}$ \\ \hline
     $R_K$   &     LU   &    $(2.488 \pm 0.010)\cdot10^{-5}$   &    2x better \\
     $\pi^+\gamma\gamma$   &    ChPT   &$< 500$ events   & $10^5$ events \\
     $\pi^0\pi^0e^+\nu$   &   ChPT   &66000 events   &      $O(10^6)$ events \\
     $\pi^0\pi^0\mu^+\nu$   &   ChPT   &      & $O(10^5)$ events \\
\end{tabular} 
\caption{NA62 sensitivities for other rare decay channels}
\label{tab:addphy} 
\end{center}
\end{table}

\section{Installation status and prospects}

As of July 2014, the majority of the detectors is in place; the installation of the straw chambers will be completed at the beginning of August. The last LAV module will be put on the beam line during August. Gigatracker and MUV1 will be installed at the beginning of the autumn. The RICH vessel is mounted and during the summer the mirrors, photomultipliers and front-end electronics will be installed.

The new readout electronics of the LKr calorimeter is ready for commissioning tests with real calibration signals. The trigger system for the calorimeter will be installed during the summer, together with the readout electronics for all the detectors and the L0 trigger processor. The PC farm is already operational.

From mid October to mid December 2014, the completed detector will be operated with beam, at the beginning to complete the commissioning of all its components and later for a first physics run to collect data at a lower intensity to reach the Standard Model sensitivity.

From 2015 to the next long accelerator shutdown, the collaboration will collect the statistics to reach the goal of the proposal.

\section{Conclusions}
The NA62 Collaboration is commissioning the detector to measure the branching ratio of the ultra-rare decay
$\kpnn$ with a precision of 10\%. In Autumn 2014 a first run with the SPS beam will collect enough data to reach the Standard Model sensitivity. The complete dataset for the measurement will be collected from 2015 onwards. The high rate and the powerful trigger will allow to collect in parallel data from other rare decays, to reach an unprecedented sensitivity.


\end{document}